\documentclass[fleqn,usenatbib,usedcolumn]{mnras}

\usepackage[T1]{fontenc}
\usepackage{ae,aecompl}

\usepackage{graphicx}	
\usepackage{amsmath}	
\usepackage{amssymb}	

\title[Inclination-type resonance in two-planet systems]{Inclination-type resonance in two-planet systems}

\author[Sotiriadis \& Libert]{
Sotiris Sotiriadis,$^{1}$
Anne-Sophie Libert,$^{1}$\thanks{E-mail: anne-sophie.libert@unamur.be}
\\
$^{1}$naXys, Department of Mathematics, University of Namur, 8 Rempart de la Vierge, 5000 Namur, Belgium
							 }
\date{Accepted XXX. Received YYY; in original form ZZZ}

\pubyear{2019}

\begin{document}
\label{firstpage}
\pagerange{\pageref{firstpage}--\pageref{lastpage}}
\maketitle

\begin{abstract}
We investigate the inclination-growth mechanisms for two-planet systems during the late protoplanetary disc phase. In previous works, much attention has been directed to the inclination-type resonance, and it has been shown that it asks for high eccentricities to be acquired during the migration of the giant planets. By adopting eccentricity and inclination damping formulae based on hydrodynamical simulations (instead of the $K$-prescription), we have carried out 20000 numerical simulations, where we vary the initial planetary eccentricities, the migration rate, and the dispersal time of the gas disc. Our results confirm that highly mutually inclined systems are unlikely to be produced by an inclination-type resonance of two migrating giant planets. However, in $\sim 1~\%$ of the simulations, inclination-type resonance is observed, and a dynamical study of the evolutions reveals that the inclination-type resonance mechanism operates in three cases: i) when the inner planet reaches the inner cavity of the disc, ii) at moderate to high eccentricities for faster migration rates, and iii) at low to moderate eccentricities after a phase of orbital destabilization and re-arrangement.     
\end{abstract}


\section{Introduction} \label{sec1}
The diversity in the orbits of extrasolar planets can be generated by planet-planet interactions during the migration in the protoplanetary disc (e.g., \citet{Moorhead2005,2008ApJ...688.1361M,Marzari2010,Matsumura2010,Libert2011a,Moeckel2012}). When the migrating planets are captured in mean-motion resonance (MMR), a growth of their eccentricities is produced, and the system can also possibly enters an inclination-type resonance (whose resonant angle is a combination of the mean longitudes and the longitudes of the ascending nodes), which induces a rapid increase of the inclinations. The inclination-type resonance was discussed by \citet{Thommes2003} for the 2/1 MMR in two-planet systems and by \citet{Libert2009} for captures into other MMRs (e.g., 3/1, 4/1 and 5/1). Both studies concluded that an inclination-type resonance is observed in two-planet systems as long as the inner planet is not very massive, one of the planets acquires high eccentricity ($e > 0.4$) and the eccentricity damping is not very efficient. In the two studies, inclination growth were observed only for $\tau_{e}/\tau_{II} > 0.2$, where $\tau_{e}$ is the timescale for eccentricity damping and $\tau_{II}$ is the timescale for \mbox{Type-II} migration.
	
More recently, \citet{Teyssandier2014} investigated, both analytically and numerically, the conditions for the onset of an inclination-type resonance for two migrating giant planets trapped in 2:1 MMR. They confirmed that, for a less massive inner planet, the system cannot enter an inclination-type resonance if the ratio $\tau_{e}/\tau_{II}$ is smaller than 0.2, even if only the eccentricity of the outer planet is damped. Therefore, they concluded that if eccentricity damping rate is much more faster than the migration rate in protoplanetary disc, i.e., $\tau_{e}/\tau_{II} \sim\!10^{-2}$ as currently believed, inclination-type resonances are unlikely to occur during the convergent migration of two gas giants. 
	
All the previous works made use of a $K$-prescription for the eccentricity (and inclination) damping, consisting in a first-order approximation ($\dot{e}/e = - K |\,\dot{a}/a|$). The $K$-factor to be considered is currently not well determined but usually estimated in the wide range 1--100. On the other hand, formulae for the eccentricity and inclination damping were derived from hydrodynamical simulations in \citet{Bitsch2013}. These explicit formulae, suitable for $n$-body codes, depend on the planetary eccentricity, inclination, and mass, as well as on the disc mass.

In \citet{Sotiriadis2017}, extensive $n$-body simulations of three giant planets on quasi-circular and quasi-coplanar orbits in the late stage of the disc were performed, using the damping formulae for eccentricity and inclination of \citet{Bitsch2013}. It was shown that while the semi-major axis and eccentricity distributions matched the observed ones for the detected giant planets, a significant proportion of the systems were found in a three-dimensional (3D) configuration with a mutual inclination higher than $10^{\circ}$, a large number of them being strongly affected by a MMR or even captured in a three-body inclination-type resonance during the migration phase (see e.g. \cite{Libert2011b} for an analytical study of the three-body inclination-type resonance).	
	
Let us note that \cite{Libert2018} have recently shown that the establishment of an inclination-type resonance can also occur at small to moderate eccentricities in two-planet systems that have previously experienced an ejection or a merging and are near a resonant commensurability at the end of the gas phase. The inclination excitation is produced by a (temporary) capture in an inclination-type resonance (associated with a vertical critical orbit along a planar elliptic family of periodic orbits), and the proximity to a spatial periodic orbit can maintain the system in a 3D configuration for a long period of time.

The goal of the present work is to study the conditions for the onset of an inclination-type resonance in two-planet systems during the disc phase, pursuing the efforts of  \citet{Thommes2003}, \citet{Libert2009} and \citet{Teyssandier2014}, but making use of the eccentricity and inclination damping formulae of \citet{Bitsch2013} instead of the simplistic $K$-prescription. As stated above, a variety of dynamical evolutions were reported in $n$-body simulations of three planets when adopting these damping formulae. In doing so, the present study extents the previous works of \citet{Thommes2003} and \citet{Libert2009} by also considering the influence of inclination damping on the establishment of the inclination-type resonance. Moreover, unlike \citet{Teyssandier2014}, the study presented here will not be restricted to the 2:1 MMR only.

The set-up of our numerical explorations is described in Section$~$\ref{section2p4}. In Section$~$\ref{section3p4}, we present several typical evolutions of two-planet systems during the disc phase, and carefully describe the inclination-growth mechanisms that operate to form the 3D system architectures observed in the simulations. An analysis of the occurrence of the different mechanisms is presented in Section$~$\ref{section4p4}. Finally, a comparison with previous studies is provided in Section$~$\ref{Conclusionsp4}, where our results are summarized. 


\begin{table}
\begin{center}
  \caption{Parameters of the seven sets of simulations. For each set, the second column describes the main features of the set, the third one the number of simulations, the fourth one the lifetime of the disc and the last column the integration time of the simulations. See text for further details.} \label{tab:table_p4} 
  \begin{tabular}{r  l  r  r  r }
   \hline
 Set& Parameters        						 & $N_{Sys}$ &$T_{disc}$ $[yr]$ & $t_{stop}$ $[yr]$      \\ \hline
 1 & Standard set      				     & 11000   & $\sim\!1\times10^6$     & $1.5\times10^6$        \\ 
 2 & $e_{0}\in[10^{-3},0.5]$          & 1500   & $\sim\!1\times10^6$     & $1.5\times10^6$        \\ 
 3 & $2~T_{disc}$       			     & 1500   & $\sim\!2\times10^6$     & $3.5\times10^6$        \\ 
 4 & $T_{disc}/2$        		     & 1500   & $\sim\!0.5\times10^6$   & $1.5\times10^6$        \\ 
 5 & $2~T_{disc}$ and $10~\tau_{II}$  & 1500   & $\sim\!2\times10^6$     & $3.5\times10^6$        \\ 
 6 & $T_{disc}/2$ and $\tau_{II}/5$   & 1500   & $\sim\!0.5\times10^6$   & $1.5\times10^6$        \\ 
 7 & $T_{disc}/2$ and $\tau_{II}/10$  & 1500   & $\sim\!0.5\times10^6$   & $1.5\times10^6$        \\ 
  \hline  
    \end{tabular}
\end{center}
\end{table}

\section{Set-up of the n-body simulations} \label{section2p4}
	In this study we follow the orbital evolution of two-planet systems and adopt the same approach as in \citet{Sotiriadis2017}, combining the speed and efficiency of a $n$-body integrator with a symplectic scheme (SyMBA code), and the improved modelization of the gas effect promoted by the hydrodynamical simulations of \citet{Bitsch2013}. Our simulations include two fully formed giant planets that evolve around a Solar-type star in the late stage of the protoplanetary disc phase. We consider that there is enough gas in the disc to affect the orbital eccentricity and inclination of both planets, while the inward migration is applied to the outer planet only.

	We have realized 20000 numerical simulations, exploring a variety of initial conditions for the planetary orbital parameters and planetary mass ratios, as well as disc masses. Seven different sets of simulations have been performed, and are presented in Table \ref{tab:table_p4}. In each ensemble, six initial disc mass values, namely 1, 2, 4, 8, 16 and 32 $M_{\rm Jup}$, were considered with equal proportion. The numerical parameters for the viscosity and the structure of the gas disc are set to the classical values $\alpha=0.005$ for the Shakura-Sunyaev parameter and $h=0.05$ for the disc aspect ratio. The initial surface density profile is $\Sigma \propto r^{-0.5} $ and the disc's inner and outer edges are set to $R_{\rm in}=0.05$ AU and $R_{\rm out}=30$ AU. A smooth transition in the gas-free inner cavity is applied with the hyperbolic tangent function $\tanh{\frac{(r - R_{\rm in})}{\Delta r}}$, where $\Delta r=0.001$ AU \citep{Matsumoto2012}.

The first set of simulations (Set 1) is referred to as the \textit{standard set}. For this set, the dispersal time for the gas is fixed to $\sim\! 10^6$ yr and we neglect the planet-disc interactions when $dM_{\rm Disc}/dt < 10^{-9}$$~$$M_{\rm star}/yr $. The rate for Type-II migration is given by the equation:
\begin{equation}\label{tII}
          \tau_{II} = \frac{2}{3} \alpha^{-1} h^{-2} \Omega_{pl}^{-1} \times \max\left(1,\frac{M_{p}}{(4\pi/3)\Sigma(r_{\rm pl})r_{\rm pl}^{2}}\right), 
\end{equation}
where $\Omega_{pl}$ is the orbital frequency of the planet and $(4\pi/3)\Sigma(r_{\rm pl})r_{\rm pl}^{2}$ the local mass of the disc,
thus the robustness of the mechanism is defined by the local mass of the disc in the vicinity of the planet. For the purpose of studying as many different MMR captures as possible, the initial semi-major axes for the inner and the outer planet are chosen randomly from the intervals $a_1 \in [3,5]$ AU and $a_2 \in [8,16]$ AU, respectively. We assume that the random initial eccentricities and inclinations follow a uniform distribution in the ranges $[0.001,0.01]$ and $[0.01^{\circ},0.1^{\circ}]$, and the same stands for the initial phase angles in the range $[0.001^{\circ},360^{\circ}]$.$~$Planetary masses are chosen also randomly from a log-uniform distribution in the interval $[0.65,5]$~$M_{\rm Jup}$.

	 In the second set of simulations (Set 2), we examine whether planets initially on non-circular orbits result in different 3D configurations after the gas phase, compared to the standard set. Thus, we keep the same modelization as in the first set, except that the initial eccentricities of the bodies are chosen randomly in the range $[0.001,0.05]$. The effect of the lifetime of the protoplanetary disc is studied in Set 3 and Set 4, where the dispersal time of the disc is fixed to $\sim\!2\times10^6$ yr and \mbox{$\sim\!0.5\times10^6$ yr}, respectively. In the last three sets, we investigate the impact of different Type-II rates on the final orbital architectures of the planetary systems. Set 5 is identical to Set 3, apart from the migration rate which is now 10 times slower. In Set 6 and Set 7, we assume that the lifetime of the disc is reduced by a factor of 2, and the migration timescales are 5 and 10 times smaller, respectively. For all the different sets, we use an integration timestep of $dt=0.001$ yr, and the systems are followed until $1.5\times10^6$ yr, except for Set 3 and Set 5 where the simulations are carried out up to $3.5\times10^6$ yr.
 \begin{figure}
       \centering
       \includegraphics[width=0.99\hsize]{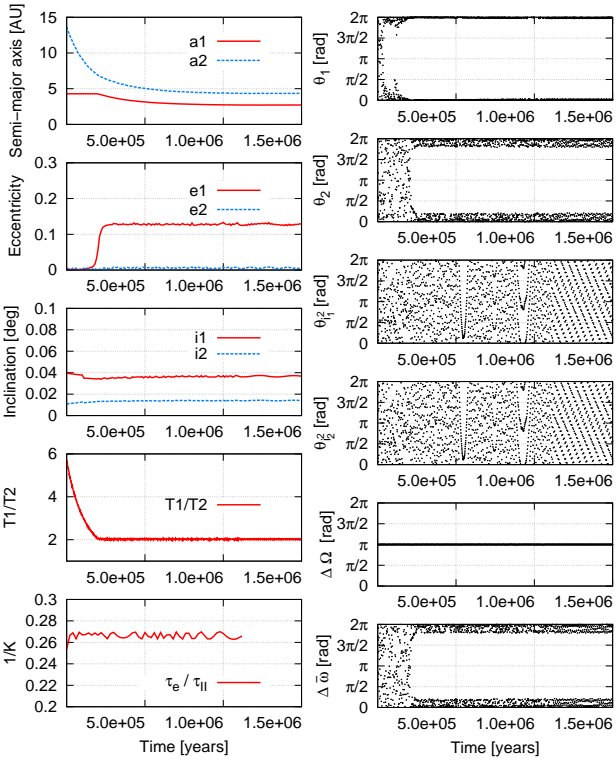} 
       \caption{The most common outcome. The giant planets emerge from the gas phase in a resonant coplanar configuration. Eccentricity damping is faster than the migration rate for the outer giant planet and $K={\tau_{e}}/{\tau_{II}}$ ratio remains roughly constant throughout the evolution of the system (bottom left panel). The planetary masses are $m_1 = 1.17$~$M_{\rm Jup}$ and $m_2 = 2.4$~$M_{\rm Jup}$. Simulation from Set 1.}
       \label{fig5.1}
 \end{figure}

 \begin{figure}
       \centering
       \includegraphics[width=0.99\hsize]{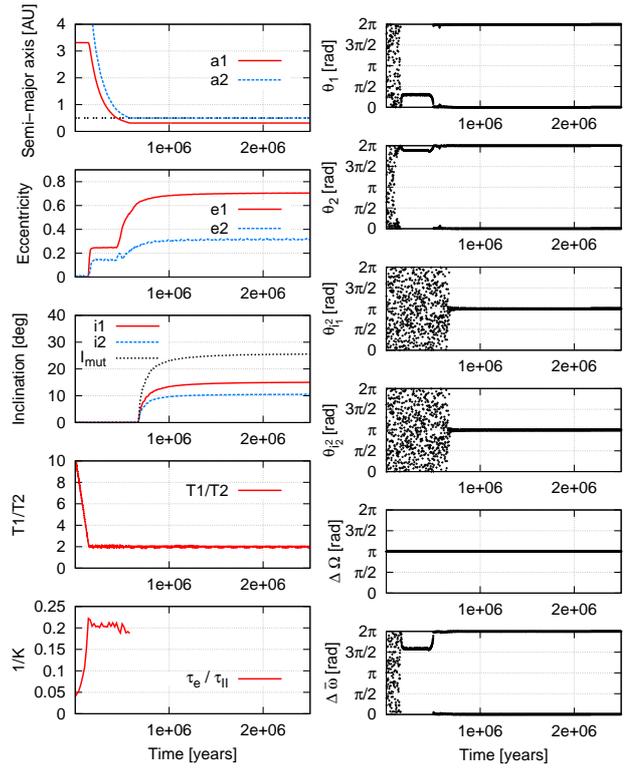}
       \caption{Formation of a 3D system by an inclination-type resonance occuring just after that the planet closer to the star reaches the inner cavity of the disc. The inner edge of the disc at 0.5 AU is indicated with the dashed black line. The planetary masses are $m_1 = 2.66$~$M_{\rm Jup}$ and $m_2 = 2.23$~$M_{\rm Jup}$, and the initial mass of the disc is 32~$M_{\rm Jup}$ (Set 3).}
       \label{fig5.2}
 \end{figure}
 \begin{figure}
       \centering
       \includegraphics[width=0.99\hsize]{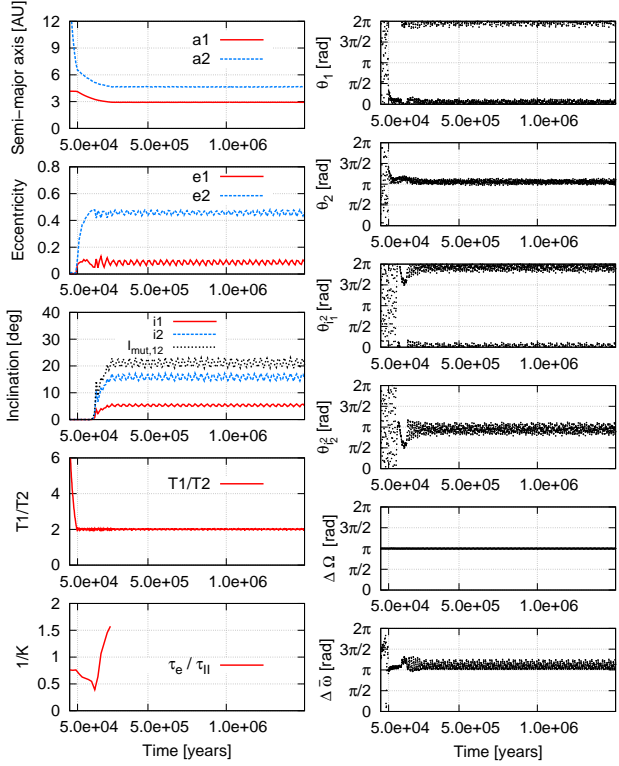}
       \caption{Formation of a 3D system by inclination-type resonance with ten times faster migration rate (Set 7). The planetary masses are $m_1 = 2.23$~$M_{\rm Jup}$ and $m_2 = 0.68$~$M_{\rm Jup}$.}
       \label{fig5.3}
 \end{figure}

 \begin{figure}
       \centering
       \includegraphics[width=0.97\hsize]{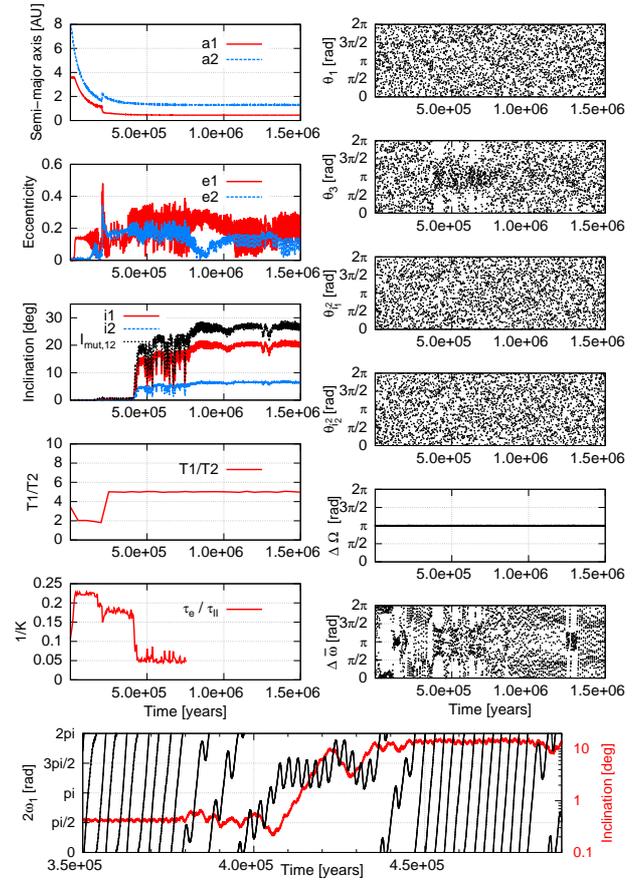}
       \caption{Formation of a 3D system by an inclination-type resonance at low to moderate eccentricities. The gas is dispersed quite rapidly (Set 4) and when inclination excitation occurs, the damping effects are not efficient enough to maintain the planets in coplanar orbits. The planetary masses are $m_1 = 1.25$~$M_{\rm Jup}$ and $m_2 = 2.20$~$M_{\rm Jup}$.}
       \label{fig5.4}
 \end{figure}

\section{Dynamical evolutions} \label{section3p4}
	In this section we present four typical evolutions observed in the simulations, and focus on the possible dynamical mechanisms producing  inclination increase in migrating two-planet systems.

	Fig.~\ref{fig5.1} illustrates the most common outcome of our simulations, where the two giant planets emerge from the gas phase in a stable and coplanar resonant configuration. The system belongs to the standard set (Set 1), with an initial disc mass of 8 $M_{\rm Jup}$. The planetary masses are $m_1 = 1.17$~$M_{\rm Jup}$ and $m_2 = 2.4$$~$$M_{\rm Jup}$, thus the mass ratio is $q\sim\!0.5$. As the outer planet migrates, the system enters the 2:1 MMR at approximately $2\times10^5$ yr and stays deep into the resonance until the end of the integration. The two resonant angles, $\theta_1=\lambda_1-2\lambda_2+\varpi_1$ and $\theta_2=\lambda_1-2\lambda_2+\varpi_2$, start to librate just after the resonant capture, as shown by the top panels of Fig.~\ref{fig5.1} (right column). Whereas the eccentricity of the inner less massive body increases and reaches the moderate value $\sim\!0.12$, the outer planet remains on a \mbox{quasi-circular} orbit. Although the system is captured in a low-order MMR, the eccentricity values reached during the evolution will remain low due to the efficiency of the damping exerted by the gas disc. The bottom panel of the left column shows the evolution of ${\tau_{e}}/{\tau_{II}}$, the ratio of the eccentricity damping timescale and the Type-II timescale, for the outer giant planet. We see that the damping on eccentricity induced by the disc is strong since ${\tau_{e}}$ is approximately four times less than the migration timescale throughout the disc's lifetime (i.e., $K\sim\!4$). In this case, no inclination-type resonance is observed (see the evolution of the inclination-type resonant angles in the right column of Fig.~\ref{fig5.1}).

	 In the next examples, we illustrate the formation of 3D systems through inclination-type resonance. Fig.~\ref{fig5.2} shows a simulation from Set 3, where an inclination-type resonance gets underway due to the presence of a planet inside of the inner cavity of the disc. While the massive disc (32 $M_{\rm Jup}$) drives the planets ($q\simeq1.15$) rapidly towards the central star, the system enters very early the 2:1 MMR, and the two angles $\theta_1=\lambda_1-2\lambda_2+\varpi_1$ and $\theta_2=\lambda_1-2\lambda_2+\varpi_2$ start to librate about $0^{\circ}$ (top right panels of Fig.~\ref{fig5.2}). However, the efficient eccentricity damping ($K\sim5$, see the bottom left panel of Fig.~\ref{fig5.2}) does not allow the bodies to acquire very eccentric orbits. Once the giant planet closer to the star reaches the disc's inner cavity, the interaction with the gas stops and its eccentricity starts to increase more. After $\sim\!5\times10^5$ yr, the outer planet also reaches the inner edge of the disc where its orbit is unaffected by the disc. The eccentricities of both bodies are rapidly high enough to trigger a capture into an inclination-type resonance. The two inclination-type resonant angles, $\theta_{i_{1}^2}=2\lambda_1-4\lambda_2+2\Omega_1$ and $\theta_{i_{2}^2}=2\lambda_1-4\lambda_2+2\Omega_2$, librate about $180^{\circ}$, and at  $\sim 2\times10^6$ yr the planets reach their maximum inclinations, both above $10^{\circ}$. The bodies continue to evolve in a stable and non-coplanar resonant configuration until the end of the simulation.

	Non-coplanar systems are also formed in Set 6 and Set~7, where we artificially decrease the Type-II migration timescale by a factor of 5 and 10, respectively. An example with $\tau_{II}/10$ is presented in Fig.$~$\ref{fig5.3} (Set 7). The outer planet, initially at $a_2=15.3$ AU, migrates rapidly inwards and, just before $5\times10^4$ yr, the 2:1 MMR is established with $\theta_1=\lambda_1-2\lambda_2+\varpi_1$ and $\theta_2=\lambda_1-2\lambda_2+\varpi_2$  librating around $0^{\circ}$ and $180^{\circ}$, respectively (top right panels of Fig.$~$\ref{fig5.3}). When the eccentricity of the less massive outer giant planet ($m_1 = 2.23 $ and $m_2 = 0.68$~$M_{\rm Jup}$) exceeds $\sim 0.4$, the system enters an inclination-type resonance, as $\theta_{i_{1}^2}=2\lambda_1-4\lambda_2+2\Omega_1$ and $\theta_{i_{2}^2}=2\lambda_1-4\lambda_2+2\Omega_2$ librate about $0^{\circ}$ and $180^{\circ}$, respectively. Let us note that the ratio ${\tau_{e}}/{\tau_{II}}$ is kept above $\sim 0.5$ throughout the lifetime of the disc (i.e., $K<2$, see bottom left panel of Fig.$~$\ref{fig5.3}). The final orbits of the planets are eventually mutually inclined with $I_{mut} > 20^{\circ}$.

	Finally, in Fig.~\ref{fig5.4}, we show an example of the establishment of an inclination-type resonance at low to moderate eccentricities. The system belongs to Set~4, wherein the dispersal time of the disc is reduced by a factor of 2. The initial mass of the disc is 32~$M_{\rm Jup}$ and the planetary masses are $m_1 = 1.25$~$M_{\rm Jup}$ and $m_2 = 2.20$~$M_{\rm Jup}$. The planets are driven rapidly to the 2:1 MMR, but the system is destabilized at $\sim 2 \times 10^5$ yr and orbital re-arrangement occurs. Rapidly the system enters the 5:1 MMR (see the libration of the resonant angle $\theta_3 = \lambda_1 -5\lambda_2 +2 \varpi_1 +2 \varpi_2$ in the second right panel), and subsequent inclination growth occurs at $\sim 4 \times 10^5$ yr due to an inclination-type resonance (at low to moderate eccentricities),  although no clear libration of the inclination-type resonant angles can be seen in Fig.$~$\ref{fig5.4} (right column). To describe more deeply the mechanism, we display, in the bottom panel of Fig.~\ref{fig5.4}, the moving average every $3\times 10^3$~yr of the angle $2 \omega_1$ (black curve) and the moving average of the inclination of the inner planet $i_1$ (red curve, in logarithmic scale), during the period of inclination growth. A correlation between the secular libration of $2 \omega_1$ and the inclination increase is clearly visible. Since $\theta_{i_3^2}=\theta_3-2\omega_1$, an inclination-type resonance can be seen, in the planetary case, as a Lidov-Kozai resonance \citep{LIDOV1962719,Kozai} embedded in a MMR. Thus, the libration of the inclination-type resonant angles  is hidden, in Fig.$~$\ref{fig5.4}, by the short period oscillations. This mechanism was previously described in \cite{Libert2018}, to which we refer for more details. The planets eventually come out from the disc phase with $I_{\rm mut} \sim 30^\circ$ and eccentricities close to 0.25.
	
	In the next section, we discuss the occurence of the dynamical mechanisms hereabove described.

 \begin{figure}
       \hspace{0.15cm}\includegraphics[width=0.93\hsize]{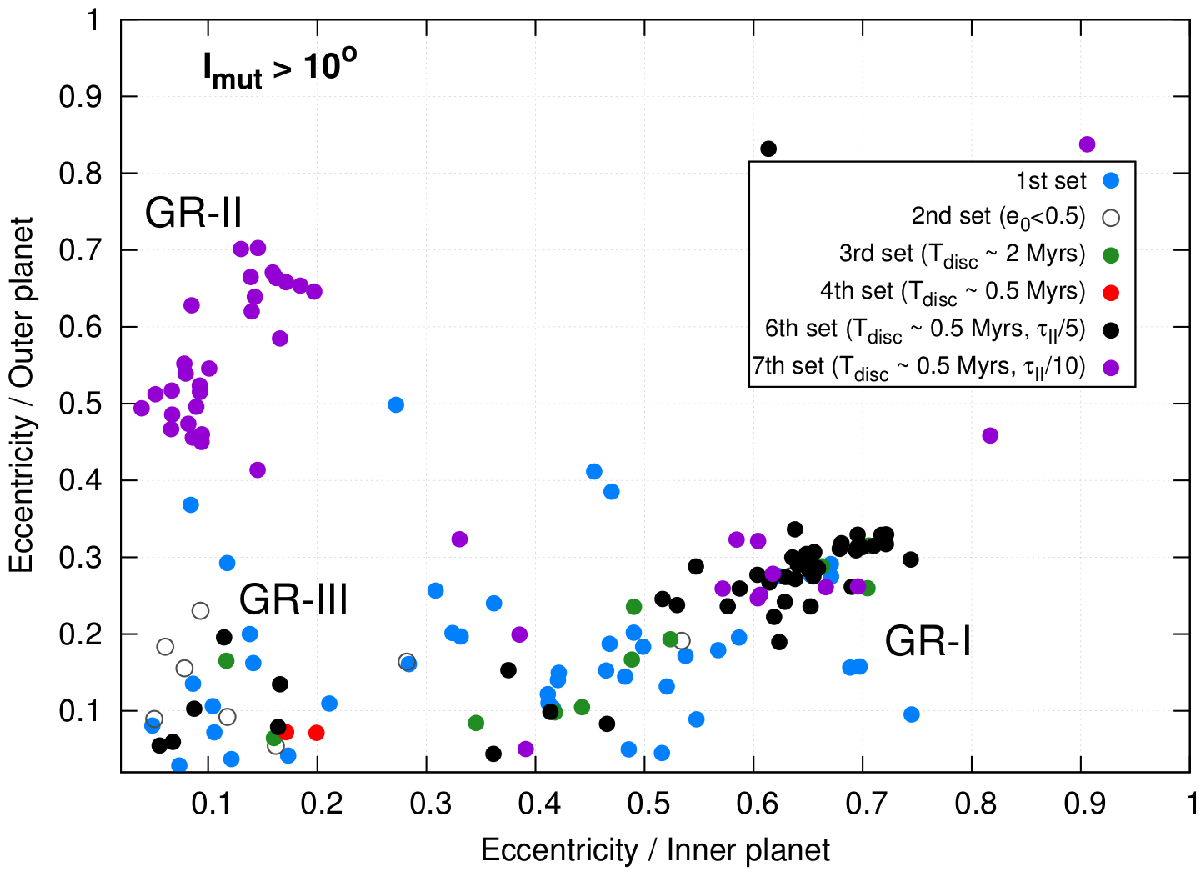}
       \includegraphics[width=1.1\hsize]{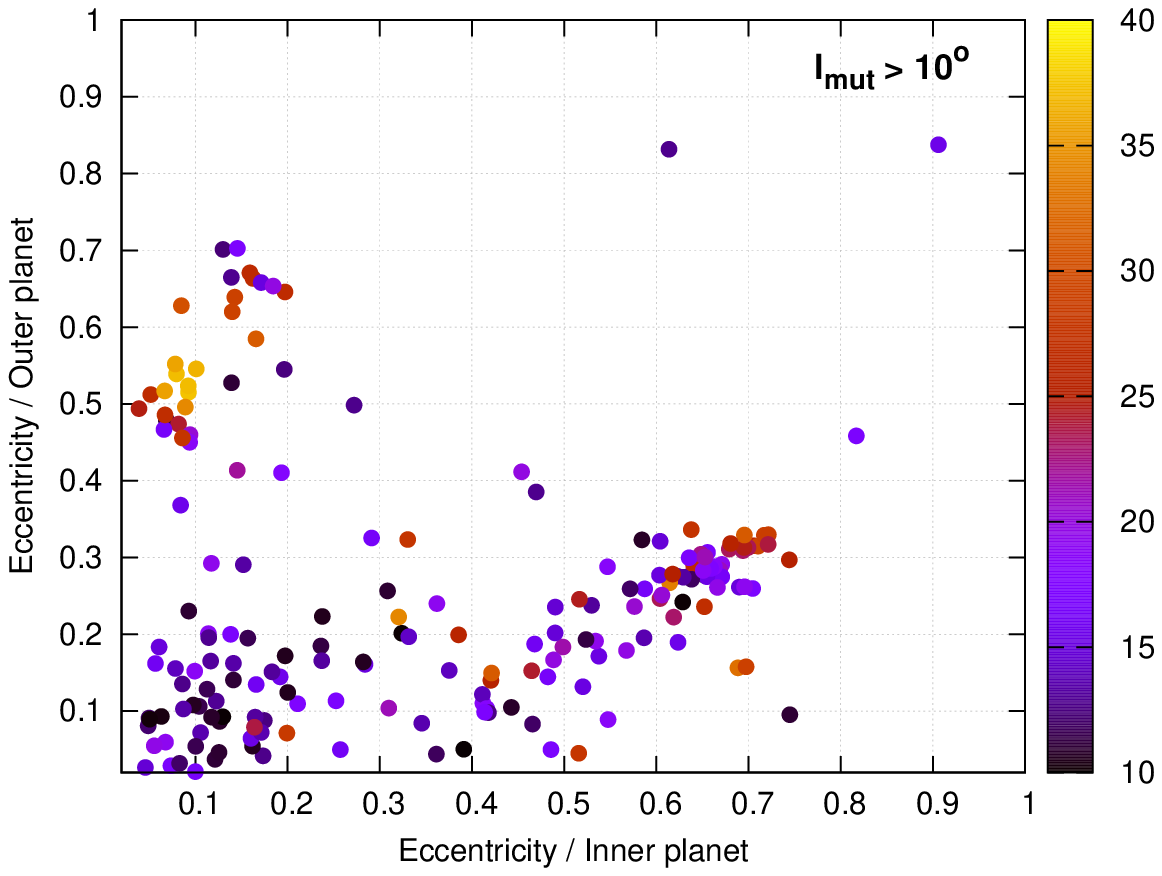}
       \caption{Eccentricities of the 153 two-planet systems with $I_{mut} > 10^{\circ}$ at the end of the simulation. In the top panel, the color code refers to the different sets, while in the bottom panel it indicates the mutual inclination value at the end of the simulation. Three groups of systems are clearly visible: GR-I at high eccentricity of the inner planet, GR-II at high eccentricity of the outer planet and GR-III at low to moderate eccentricities for both planets. See text for a detailed description.}
       \label{fig5.8}
 \end{figure}

\section{Orbital configurations after the dispersal of the disc}\label{section4p4}

	In this section, we focus on the orbital characteristics of the systems observed in the simulations with final $I_{mut}$ higher than $10^{\circ}$, and investigate the different formation mechanisms leading to the inclination excitation. 

	The vast majority of the systems formed in the simulations end up in a coplanar configuration as a result of the strong eccentricity and inclination damping of the gas disc. Nonetheless, as shown in the previous section, instabilities and MMR captures can lead the migrating planets away from the disc midplane. Thereafter, the inclined bodies either return to coplanar orbits due to the strong damping, or maintain their 3D configuration until the end of the protoplanetary disc phase.

\begin{table}
\centering
  \caption{Frequency of the different inclination excitation mechanisms producing the highly mutually inclined systems ($I_{mut}>10^{\circ}$), for each set of simulations.} \label{tab:table2_p4}

 \begin{tabular}{r c c c c c c c}
  \hline 
  
  &  Set 1   	& Set 2  & Set 3 	&  Set 4	  	&Set 5 & Set 6 & Set 7   	\\ 
 Number	  & 11000	& 	1500     & 1500 & 1500 & 1500 & 1500  & 1500				\\ 
 GR-I &  30	& 	3   & 9 & 0 & 0 & 38  & 11 \\   
  GR-II &  0	& 	0   & 0 & 0 & 0 & 0  & 28 \\   
  GR-III &  12	& 	6   & 2 & 2 & 0 & 6  & 0 \\   
 {\scriptsize Other/Instab} &   3	& 	0   & 0 & 0 & 0 & 1  & 2 \\   
	  
 	  \hline

 \end{tabular}
 
\end{table}

	More specifically, the huge majority of the 19413 systems that consist of the two planets at the end of simulation\footnote{In the simulations, 587 planetary systems suffer from merging or ejection, especially the systems issued from simulations with reduced migration timescales (Set 6 and Set~7).}, are found in a coplanar configuration. However, for $\sim 5\%$ of the simulations (1050 systems), the giant planets are found with mutually inclined orbits ($I_{mut}\!>$~$1^{\circ}$), while significantly mutually inclined architectures with $I_{mut} > 10^{\circ}$ represent $\sim 1\%$ of the simulations (153 systems). 
	Hereafter, we focus on the analysis of the latter category of systems and try to perceive the different formation scenarios.

	The final eccentricities of the 153 highly mutually inclined systems are examined in Fig.$~$\ref{fig5.8}. The color code gives additional information on the different sets in the top panel and the final mutual inclinations in the bottom panel. We clearly observe three distinct regions of systems in the $(e_1,e_2)$ plane. The first region, denoted GR-I, covers the area for $e_{inner}>~0.25$ and $e_{outer}<~0.35$. It mainly consists of systems that have been captured in an inclination-type resonance (at moderate to high eccentricities), when the inner planet is interior to the inner edge of the disc (Set 6), similarly to the evolution shown in Fig.$~$\ref{fig5.2}. The mean mass ratio $<q>$ for this region is $\sim$~1. 

	The second region, GR-II, is found at $e_{inner}<~0.2$ and $e_{outer}>~0.4$.  It only consists of systems from the simulations with ten times faster Type-II migration rate, for which the inclination-type resonance occurs at moderate to high eccentricities during the evolution in the disc. In this region, $<q>\sim$~3, i.e., the inner planet is always more massive than the outer one, which explains the higher value of the eccentricity of the outer planet. The evolutionary path of a system that belongs to this group has been displayed in Fig.$~$\ref{fig5.3}.

	The GR-III region consists of 3D systems that have acquired their inclinations through an inclination-type resonance at low to moderate eccentricities (see the example of Fig.$~$\ref{fig5.4}), after a phase of destabilization of the system and subsequent orbital re-arrangement.

 \begin{figure}
       \centering
       \includegraphics[width=0.95\hsize]{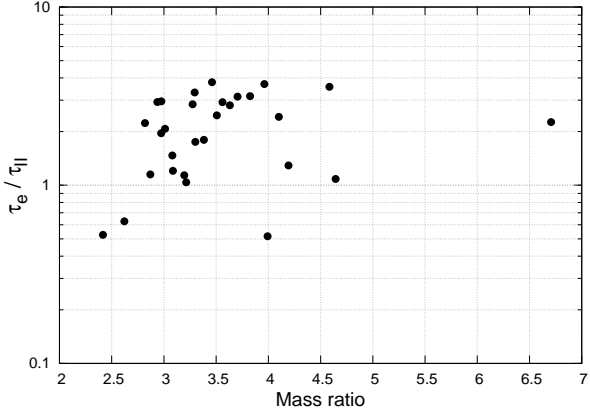}
       \caption{Ratio of the eccentricity damping timescale over the migration timescale ${\tau_{e}}/{\tau_{II}}$ vs. the mass ratio, for the inclination-type resonant systems of GR-II. The ratio is evaluated for the outer planet, just before the end of the interaction with the disc.}
       \label{fig5.11}
 \end{figure}

Let us note that a large majority of the 3D planetary systems are found in the 2:1 MMR. It is especially the case for the systems from Set 6 and Set 7, since a fast migration makes difficult a capture in a high-order MMR (see e.g. \citet{Libert2009}). For these systems, higher eccentricity and mutual inclination values (see the second panel of Fig.$~$\ref{fig5.8}) are reached during the evolution in the gas disc. However, as observed in our simulations, an inclination-type resonance can also possibly follow a capture in 3:1, 4:1, 5:1 or 6:1 MMR (see for instance the evolution in Fig.~\ref{fig5.4}).

Table \ref{tab:table2_p4} shows the frequency of the above-mentioned formation mechanisms, for each set of simulations. We observe that around $59\%$ of the highly mutually inclined systems acquire their 3D configuration by an inclination-type resonance with planets inside of the inner cavity of the disc (GR-I). They principally originate from the standard set as well as Set~6 and Set~7 (faster migration rates). Around a fifth of the systems belong to GR-II characterized by inclination-type resonance captures associated with a fast migration rate, and they are all issued from Set~7. Libration of the inclination-type resonant angles at low to moderate eccentricities (GR-III) is observed in $18\%$ of the systems. The GR-III systems emanate from nearly all the simulation sets, but to a large extent from the standard set. The remaining $4\%$ of simulations generate inclination increase due to orbital instability (no inclination-type resonance observed).

According to Table \ref{tab:table2_p4}, about $0.5\%$ of the planetary systems formed in the standard set (Set 1) are highly mutually inclined, and $\sim25\%$ of them do experience an inclination-type resonance at low to moderate eccentricities. While modifications in the initial planetary eccentricies (Set 2) and the lifetime of the disc (Set 3, Set 4 and Set 5) do not significantly increase the number of systems formed in highly mutually inclined configuration, faster migration rates clearly boost inclination excitation. Approximately $3\%$ of the planetary systems are found highly non-coplanar in Set 6 and Set 7. 
	
\section{Discussion and conclusions} \label{Conclusionsp4}

In this work, we investigated the inclination-growth mechanisms for systems with two giant planets during the late protoplanetary disc phase. We performed 20000 simulations considering Type-II migration and the eccentricity and inclination damping modelling of \citet{Bitsch2013}. Besides adopting different initial orbital parameters of the planets and disc masses, seven sets of simulations were carried out with different disc lifetimes and migration rates. Let us recall that the migration of the planets in Type-II regime due to angular momentum exchange with the disc is on a similar timescale as the viscous accretion time. However, the actual Type-II migration rate is still under debate and may depart from the viscous time \citep{Has2013,Duffell2015,Durmann2015,Kanagawa2018,Scardoni2020}. In particular, the prescription for the viscosity depends on the $\alpha$ parameter whose appropriate value is unknown. Considering different migration rates includes this uncertainty.
	
The vast majority of the two-planet systems formed in our simulations end up in coplanar orbits, as inclination-type resonance capture is rarely observed. This is in agreement with previous works. In particular, \citet{Thommes2003} found that inclination-type resonance cannot be established for a planetary system in the 2:1 MMR if $K>5$ and the mass ratio $q<2$, since, due to the strong damping, the planetary eccentricities do not reach sufficiently high values for the inclination-type resonance to be established.  \citet{Lee20091662} showed that an inclination-type resonance capture is also possible for $q>2$ if the Type-II migration rate is slower than the one adopted by \citet{Thommes2003}. This result was extended for other MMRs by \citet{Libert2009}, showing that inclination-type resonance is only observed when the inner planet is not very massive, one of the planets acquires high eccentricity ($e > 0.4$) and the eccentricity damping is not too efficient ($K<5$). Regarding the 2:1 MMR, \citet{Teyssandier2014} confirmed that, for a less massive inner planet, inclination-type resonance only occurs if  ${\tau_{e}}/{\tau_{II}} > 0.2$ (i.e., $K<5$). All these works conclude that, given the current estimation of the $K$-factor to 100, inclination-type resonance capture is unlikely to occur in protoplanetary discs. 

	Using a different approach for the eccentricity and inclination damping based on hydrodynamical simulations, we have found that the evolution presented in Fig.$~$\ref{fig5.1} showing a planetary system remaining coplanar during the protoplanetary disc phase is also the typical outcome of our simulations. For comparison with previous studies using the $K$-factor, our simulations are characterized by the mean ratio $<{\tau_{e}}/{\tau_{II}}> \sim\! 0.12$. The strong eccentricity damping explains that the onset of inclination-type resonance at moderate to high eccentricities is quite uncommon in our study.
	
	However, around $1\%$ of the systems formed in the simulations are non-coplanar, with a mutual inclination higher than $10^{\circ}$. We deeply studied their dynamical evolution and identified three groups of systems, depending on the values of their final eccentricities. GR-I involves almost half of the systems, and consists of the simulations where an inclination-type resonance has been triggered interior to the disc's inner edge. The reason for inclination excitation is obvious, since any interaction between the planets and the gas is halted in that region and no more damping is applied to the planets (Fig.$~$\ref{fig5.2} is a typical case from this category). 
	
	GR-II consists of the systems where the inclination-type resonance occurs inside the protoplanetary disc at moderate to high eccentricities ($\sim 20\%$ of the 3D systems, all from Set~7 with ten times faster migration rate, see example in Fig.$~$\ref{fig5.3}). To show how the ratio of the eccentricity damping timescale over the migration timescale affects the onset of the inclination-type resonance for these systems, we present, in Fig.$~$\ref{fig5.11}, the ratio ${\tau_{e}}/{\tau_{II}}$, for the outer planet, as a function of the mass ratio of the planets. We observe that the establishment of an inclination-type resonance for these systems requires ${\tau_{e}}/{\tau_{II}}>0.5$ (the mean value is 2.14 in this subset), and a more massive inner giant ($q > 2$). Whereas our simulations adopt a more complex prescription for eccentricity and inclination damping, our results are in agreement with the work of \citet{Teyssandier2014}. For this reason, no system from Set 1 (standard set) was found in this category. 
	
	However, our work highlights that inclination increase can be driven in two-planet systems by a (temporary) capture in an inclination-type resonance at low to moderate eccentricities. We reported that, for a third group of nearly $20\%$ of the 3D systems, GR-III, an inclination-type resonance at low to moderate eccentricities can excite the inclinations, especially after a phase of orbital instability and re-arrangement, as observed in Fig.$~$\ref{fig5.4}. In this case, the ratio ${\tau_{e}}/{\tau_{II}}$ can be lower than 0.5. This dynamical mechanism can also operate in three-planet systems after a phase of ejection or merging, as previously observed in \citet{Libert2018}. 
	
	Although only observed in $\sim1\%$ of the simulations, inclination-type resonance is a possible mechanism for the formation of highly non-coplanar systems. This percentage is strongly dependent on the parameters of the damping formulae for eccentricity and inclination (e.g., in our case, the disc mass considered) and on the Type-II migration timescale which is still in debate.

\section*{Acknowledgements}

This work was supported by the Fonds de la Recherche Scientifique-FNRS under Grant No. T.0029.13 (ExtraOrDynHa research project). Computational
resources have been provided by the PTCI (Consortium des \'Equipements de Calcul Intensif CECI), funded by the FNRS-FRFC, the Walloon Region, and the
University of Namur (Conventions No. 2.5020.11, GEQ U.G006.15, 1610468 et
RW/GEQ2016).


\bibliographystyle{mnras}
\bibliography{Biblio} 

\bsp	
\label{lastpage}
\end{document}